\documentclass[%
 aip,
 jmp,%
 amsmath,amssymb,
 reprint,%
]{revtex4-1}

\usepackage{graphicx}
\usepackage{dcolumn}
\usepackage{bm}
\usepackage{epstopdf}

\begin{document}

\preprint{AIP/123-QED}

\title{Strongly lensed gravitational waves as the probes to test the cosmic distance duality relation}

\author{Hai-Nan Lin}
\affiliation{Department of Physics, Chongqing University, Chongqing 401331, China}
\author{Xin Li}
 \affiliation{Department of Physics, Chongqing University, Chongqing 401331, China}
\author{Li Tang}%
 \email{tang@cqu.edu.cn}
\affiliation{Department of Math and Physics, Mianyang Normal University, Mianyang 621000, China}
\affiliation{Department of Physics, Chongqing University, Chongqing 401331, China}
 %


\begin{abstract}
The cosmic distance relation (DDR) associates the angular diameters distance ($D_A$) and luminosity distance ($D_L$) by a simple formula, i.e., $D_L=(1+z)^2D_A$. The strongly lensed gravitational waves (GWs) provide a unique way to measure $D_A$ and $D_L$ simultaneously to the GW source, hence can be used as probes to test DDR. In this paper, we prospect the use of strongly lensed GW events from the future Einstein Telescope to test DDR. We write the possible deviation of DDR as $(1+z)^2D_A/D_L=\eta(z)$, and consider two different parametrizations of $\eta(z)$, namely, $\eta_1(z)=1+\eta_0 z$ and $\eta_2(z)=1+\eta_0 z/(1+z)$. Numerical simulations show that, with about 100 strongly lensed GW events observed by ET, the parameter $\eta_0$ can be constrained at $1.3\%$ and $3\%$ levels for the first and second parametrizations, respectively.

\end{abstract}

\maketitle

\section{Introduction}\label{sec:introduction}

In astronomy and cosmology we often require to measure the distance to a celestial body in the sky far away from us. However, due to the accelerating expansion of the universe, there is no unique way to define distance. Among several definitions of distance, the luminosity distance ($D_L$) and angular diameter distance ($D_A$) are widely used. The definition of luminosity distance is based on the fact that the measured bolometric flux from a spherically symmetric radiating body is inversely proportional to the square of distance to the radiating source. The angular diameter distance, on the other hand, is defined as the ratio of transverse linear size to angular size of a celestial body. In the standard cosmology, the spacetime is governed by Einstein's general relativity, and these two distances are correlated by Etherington's distance duality relation (DDR), i.e., $D_L(z)=(1+z)^2D_A(z)$ \cite{Etherington:1933,Etherington:2007}. The validity of DDR requires that photons propagate along null geodesics and the number of photons is conserved \cite{Ellis:1971,Ellis:2007}. The violation of DDR may be caused by e.g. the extinction of photon by intergalactic dust \cite{Corasaniti:2016}, the coupling of photon with other particles \cite{Bassett:2003vu}, the variation of fundamental constants \cite{Ellis:2013}, and so on. DDR is a fundamental relation in the standard cosmological model, hence testing its validity is of great importance.

Several methods have been proposed to test DDR, see e.g. Refs.\cite{Holanda:2010vb,Piorkowska:2011nhd,Li:2011mdk,Yang:2013coa,Liang:2013yst,Costa:2015lja,Holanda:2016msr,Liao:2016uzb,Ma:2016bjt,Holanda:2016zpz,Li:2018,Hu:2018yah,Lin:2018mdj,Liao:2019xug}. In order to test DDR, one needs to independently measure both $D_L$ and $D_A$ at the same redshift $z$. The measurement of luminosity distance is relatively easy. For example, as the standard candles, type-Ia supernovae (SNe Ia) provide an excellent tool to measure $D_L$ up to redshift $z\sim 2.3$ \cite{Scolnic:2017caz}. According to the luminosity-period relation of Cepheid variables, we can also measure $D_L$, but to a relatively lower redshift. Besides, the gravitational waves (GWs) can be used as the standard sirens to measure $D_L$ \cite{Abbott:2016ajs,Abbott:2017lpn}. There are also several methods to measure the angular diameter distance. One of the most precise way to measure $D_A$ is using the standard ruler baryonic acoustic oscillations (BAO), which can be measured up to redshift $z\sim2.34$ \cite{Delubac:2015}, comparable to the furthest SNe Ia detected at present. We can also measure $D_A$ from the Sunyaev-Zel'dovich effect of galaxy clusters \cite{Filippis:2005,Bonamente:2006ct} and the angular size of ultra-compact radio sources \cite{Jackson:2006bg}, but the uncertainty is much larger than BAO. In addition, strong gravitational lensing systems can provide information of angular diameter distance \cite{Liao:2016uzb,Liao:2019xug}. However, in the ordinary quasar lensing systems where only two images are seen, only the ratio of distances between lens to source and between observer to source can be obtained, unless the time delay between two images can be observed to break the degeneracy.

One shortcoming of the above methods is that $D_A$ and $D_L$ are measured from different sources at different redshifts, hence couldn't be directly used to test DDR. To solve this problem, we can first apply some special techniques such as interpolations \cite{Liang:2013yst} and Gaussian processes \cite{Lin:2018mdj} to reconstruct the $D_L-z$ relation, then $D_L$ can be calculated at any redshift we want. In addition, we can also use the nearest neighbourhood method \cite{Holanda:2010vb} to pick up $D_L$ and $D_A$ that are measured at approximately equal redshifts. After these procedures, $D_L$ and $D_A$ can be compared at the same redshift. However, the above methods have not considered the fact that $D_L$ and $D_A$ are usually measured at different sky directions. If DDR really holds, this is not a problem. However, if there is any violation of DDR, e.g. caused by photo extinction by intergalactic dust, then the measured $D_L$ may depend on the sky direction, since different line-of-sight directions may have different environments. Therefore, it is unreasonable to test DDR using $D_A$ and $D_L$ measured at different sky directions. The most ideal way to avoid this problem is to measure $D_A$ and $D_L$ from the same source. But this is not trivial, since it is difficult to find a source who can play the roles of standard candle and standard ruler simultaneously.

In a recent paper \cite{Lin:2020mdh}, we proposed a new method to measure $D_L$ and $D_A$ simultaneously from the strongly lensed GW events. We have shown that, if the image positions of GW source and the relative time delay between different images can be observed simultaneously, and if the redshifts of lens and source can be measured independently, then we can extract $D_L$ and $D_A$ to the GW source. Therefore, the strongly lensed GWs provide a unique way to measure the luminosity distance and angular distance simultaneously to the same source, thus can be used to test DDR. A rough estimate shows that, with $\sim 100$ such events, DDR can be constrained at several percentage level. The third generation ground-based GW detectors such as the Einstein Telescope (ET) is expected to observe hundreds of strongly lensed GW events in the future \cite{Biesiada:2014kwa,Ding:2015uha}. Therefore, it is meaningful to prospect the accuracy of the future GW observations in constraining DDR. In this paper, based on the designed sensitivity of ET, we will use numerical simulation to investigate the ability of strongly lensed GW events in constraining DDR.

The rest of this paper is arranged as follows: In section {\ref{sec:method}}, we introduce the method of how to measure angular diameter distance and luminosity distance simultaneously from the strongly lensed GW events. In section {\ref{sec:simulations}}, based on the sensitivity of ET, we use Monte Carlo simulations to investigate the accuracy of using strongly lensed GW events to constrain DDR. Finally, discussion and conclusions are given in section {\ref{sec:discussion}}.

\section{Methodology}\label{sec:method}
\subsection{Measure $D_A$ from strong lensing}\label{sec:SL}

Suppose a GW burst is strongly lensed by a foreground galaxy. For simplicity we assumed that the lens galaxy is spherically symmetric. Specifically, we  take the singular isothermal sphere (SIS) model as an example. With this configuration, we will see two images at opposite sides of the lens position. The Einstein radius $\theta_E=|\theta_1-\theta_2|/2$ is given by \cite{Mollerach:2002}
\begin{equation}\label{eq:thetaE}
  \theta_E=\frac{4\pi\sigma_{\rm SIS}^2D_A(z_l,z_s)}{c^2D_A(z_s)},
\end{equation}
where $\sigma_{\rm SIS}$ is the velocity dispersion of the lens galaxy, $\theta_1$ and $\theta_2$ are the image positions with respect to the lens galaxy, $D_A(z_s)$ and $D_A(z_l,z_s)$ are the angular diameter distances from the observer to source and from the lens to source, respectively. Inverting equation (\ref{eq:thetaE}) we can obtain the distance ratio
\begin{equation}\label{eq:Dls2Ds}
  R_A\equiv\frac{D_A(z_l,z_s)}{D_A(z_s)}=\frac{c^2\theta_E}{4\pi\sigma_{\rm SIS}^2}.
\end{equation}
If the angular resolution of the GW detector is high enough such that the angular positions of the two images ($\theta_1$ and $\theta_2$) can be well localized, and if the velocity dispersion of the lens galaxy can be measured independently, then the distance ratio $R_A$ can be calculated according to equation (\ref{eq:Dls2Ds}).

Different images propagate along different paths and feel different gravitational potentials, so have different time consumptions when arriving to the detector. The time delay between two images is given by \cite{Mollerach:2002}
\begin{equation}\label{eq:timedelay}
  \Delta t=(1+z_l)\frac{D_{\Delta t}}{c}\Delta\phi,
\end{equation}
where
\begin{equation}\label{eq:D_timedelay}
  D_{\Delta t}\equiv\frac{D_A(z_l)D_A(z_s)}{D_A(z_l,z_s)}=\frac{c}{1+z_l}\frac{\Delta t}{\Delta\phi}
\end{equation}
is the time-delay distance, and
\begin{equation}\label{eq:Fermi_potential}
  \Delta\phi=\frac{(\theta_1-\beta)^2}{2}-\Psi(\theta_1)-\frac{(\theta_2-\beta)^2}{2}+\Psi(\theta_2)
\end{equation}
is the Fermat potential difference between two paths, $\Psi(\theta)$ is the rescaled projected gravitational potential of the lens galaxy. For the singular isothermal spherical lens, $\Psi(\theta)=\theta_E|\theta|$. If the gravitational potential of the lens galaxy can be measured from photometric and spectroscopic observations, and if the time delay between two images can be recorded, then we can calculate the time-delay distance $D_{\Delta t}$ according to equation (\ref{eq:D_timedelay}), given that the spectroscopic redshift of the lens galaxy is precisely known.

In a spatially flat universe, the comoving distance is related to the angular diameter distance by $r(z_s)=(1+z_s)D_A(z_s)$, $r(z_l)=(1+z_l)D_A(z_l)$, $r(z_l,z_s)=(1+z_s)D_A(z_l,z_s)$, where the comoving distance from lens to source is simply given by $r(z_l,z_s)=r(z_s)-r(z_l)$. Therefore, the angular diameter distance from lens to source can be written as
\begin{equation}\label{eq:Dls}
  D_A(z_l,z_s)=D_A(z_s)-\frac{1+z_l}{1+z_s}D_A(z_l).
\end{equation}

From equations (\ref{eq:Dls2Ds})(\ref{eq:D_timedelay})(\ref{eq:Dls}) we can uniquely solve for $D_A(z_s)$, which reads
\begin{equation}\label{eq:DA_zs}
  D_A(z_s)=\frac{1+z_l}{1+z_s}\frac{R_AD_{\Delta t}}{1-R_A},
\end{equation}
where $R_A$ and $D_{\Delta t}$ are given by equations (\ref{eq:Dls2Ds}) and (\ref{eq:D_timedelay}), respectively. Assuming that $R_A$ and $D_{\Delta t}$ are uncorrelated, we can obtain the uncertainty on $D_A(z_s)$ using the standard error propagating formulae,
\begin{equation}\label{eq:error_DA_zs}
  \frac{\delta D_A(z_s)}{D_A(z_s)}=\sqrt{\left(\frac{\delta R_A}{R_A(1-R_A)}\right)^2+\left(\frac{\delta D_{\Delta t}}{D_{\Delta t}}\right)^2},
\end{equation}
where the uncertainty on $R_A$ propagates from the uncertainties on $\theta_E$ and $\sigma_{\rm SIS}$,
\begin{equation}\label{eq:error_RA}
  \frac{\delta R_A}{R_A}=\sqrt{\left(\frac{\delta\theta_E}{\theta_E}\right)^2+4\left(\frac{\delta\sigma_{\rm SIS}}{\sigma_{\rm SIS}}\right)^2},
\end{equation}
and the uncertainty on $D_{\Delta t}$ propagates from the uncertainties on $\Delta t$ and $\Delta\phi$,
\begin{equation}\label{eq:error_Dt}
  \frac{\delta D_{\Delta t}}{D_{\Delta t}}=\sqrt{\left(\frac{\delta \Delta t}{\Delta t}\right)^2+\left(\frac{\delta\Delta\phi}{\Delta\phi}\right)^2}.
\end{equation}
If the physical quantities ($z_l$, $z_s$, $\Delta t$, $\Delta \phi$, $\theta_E$, $\sigma_{\rm SIS}$) are measured, $D_A(z_s)$ and its uncertainty can be obtained from equations (\ref{eq:DA_zs})--(\ref{eq:error_Dt}).

\subsection{Measure $D_L$ from GW signals}\label{sec:GW}

As the standard sirens, GWs provide an excellent tool to measure the luminosity distance. The self-calibrating property of GWs makes the measurement of $D_L$ being independent of any other cosmological probes, and also independent of cosmological model. According to general relativity, GW has two polarization states, which are written as $h_{+}(t)$ and $h_{\times}(t)$. GW detectors based on the interferometers such as ET measure the change of difference of two optical paths caused by the spacetime fluctuation when GW signals pass. The response of GW detectors on GW signals depends on the spacetime strain, which is the linear combination of two polarization states,
\begin{equation}
  h(t)=F_+(\theta,\varphi,\psi)h_+(t)+F_\times(\theta,\varphi,\psi)h_\times(t),
\end{equation}
where the beam-pattern functions $F_+(\theta,\varphi,\psi)$ and $F_\times(\theta,\varphi,\psi)$ not only depend on the configuration of detector, but also depend on the position of GW source on the sky $(\theta,\varphi)$ and the polarization angle $\psi$.

The Einstein Telescope (ET) \cite{ET} is a third generation ground-based GW detector under designed. It consists of three interferometer arms of 10 kilometers length, arranged along three sides of an equilateral triangle, respectively. ET is sensitive in the frequency range $1-10^4$ Hz, and it is expected to be able to detect GW signals produced by the coalescence of compact binary system up to redshift $z\sim 5$. The beam-pattern functions for ET are given as \cite{Zhao:2011}
\begin{eqnarray}\nonumber
  F_+^{(1)}(\theta,\varphi,\psi)&=&\frac{\sqrt{3}}{2}\Big[\frac{1}{2}(1+\cos^2\theta)\cos2\varphi\cos2\psi\\\nonumber
  &&-\cos\theta\sin2\varphi\sin2\psi\Big],\\\nonumber
  F_\times^{(1)}(\theta,\varphi,\psi)&=&\frac{\sqrt{3}}{2}\Big[\frac{1}{2}(1+\cos^2\theta)\cos2\varphi\sin2\psi\\\nonumber
  &&+\cos\theta\sin2\varphi\cos2\psi\Big],\\\nonumber
  F_{+,\times}^{(2)}(\theta,\varphi,\psi)&=&F_{+,\times}^{(1)}(\theta,\varphi+2\pi/3,\psi),\\
  F_{+,\times}^{(3)}(\theta,\varphi,\psi)&=&F_{+,\times}^{(1)}(\theta,\varphi+4\pi/3,\psi).
\end{eqnarray}

In this paper, we only consider the GW signals produced by the coalescence of compact binary systems (e.g. NS-NS binary and NS-BH binary). In signal processing of GWs, it is convenient to work in the Fourier space. Using the post-Newtonian and stationary phase approximation, the spacetime strain $h(t)$ can be written in the the Fourier space by \cite{Zhao:2011,Sathyaprakash:2009}
\begin{equation}\label{eq:fourier_strain}
  \mathcal{H}(f)=\mathcal{A}f^{-7/6}\exp[i(2\pi f t_0-\pi/4+2\psi(f/2)-\varphi_{(2,0)})],
\end{equation}
where
\begin{eqnarray}\label{eq:amplidude}\nonumber
  \mathcal{A}&=&\frac{1}{D_L}\sqrt{F_+^2(1+\cos^2\iota)^2+4F_\times^2\cos^2\iota}\\
  &&\times\sqrt{\frac{5\pi}{96}}\pi^{-7/6}\mathcal{M}_c^{5/6},
\end{eqnarray}
is the Fourier amplitude, $\iota$ is the inclination of the binary's orbital plane, $D_L$ is the luminosity distance from the GW source to the detector, $\mathcal{M}_c=M\eta^{3/5}$ is the chirp mass, $M=m_1+m_2$ is the total mass, $\eta=m_1m_2/M^2$ is the symmetric mass ratio, $m_1$ and $m_2$ are the component masses of the binary in comoving frame. For a GW source at redshift $z$, $\mathcal{M}_c$ in equation (\ref{eq:amplidude}) should be interpreted as the chirp mass in observer frame, which is related to the chirp mass in comoving frame by $\mathcal{M}_{c,{\rm obs}}=(1+z)\mathcal{M}_{c,{\rm com}}$ \cite{Krolak:1987}. The exponential term on the right-hand-side of equation (\ref{eq:fourier_strain}) represents the phase of GW strain, whose explicit form can be found in Ref.\cite{Sathyaprakash:2009}, but it is unimportant in our study.

The signal-to-noise ratio (SNR) of a GW signal is given by the square root of the inner product of the spacetime strain in Fourier space, namely \cite{Sathyaprakash:2009}
\begin{equation}\label{eq:snr}
  \rho_i=\sqrt{\langle \mathcal{H},\mathcal{H}\rangle},
\end{equation}
where the inner product is defined as
\begin{equation}
  \langle a,b \rangle=4\int_{f_{\rm lower}}^{f_{\rm upper}}\frac{\tilde{a}(f)\tilde{b}^*(f)+\tilde{a}^*(f)\tilde{b}(f)}{2}\frac{df}{S_h(f)},
\end{equation}
where $\tilde{a}$ and $a^*$ represent the Fourier transformation and complex conjugation of $a$, respectively, $S_h(f)$ is the one-side noise power spectral density (PSD) characterizing the sensitivity of the detector on spacetime strain, $f_{\rm lower}$ and $f_{\rm upper}$ are the lower and upper cutoffs of the frequency. The PSD for ET is given by \cite{Mishra:2010lfdk,Cai:2017aea}
\begin{eqnarray}\nonumber
  S_h(f)&=&10^{-50}(2.39\times10^{-27}x^{-15.64}+0.349x^{-2.145}\\
  &&+1.76x^{-0.12}+0.409x^{1.1})^2 ~ {\rm Hz}^{-1}.
\end{eqnarray}
Following Ref.\cite{Zhao:2011}, we assume $f_{\rm lower}=1$ Hz and $f_{\rm upper}=2f_{\rm LSO}$, where $f_{\rm LSO}=1/(6^{3/2}2\pi M_{\rm obs})$ is the orbit frequency at the last stable orbit, $M_{\rm obs}=(1+z)(m_1+m_2)$ is the total mass in observer frame. If $N$ independent detectors form a network and detect the same GW source simultaneously, the combined SNR is given by
\begin{equation}\label{eq:snr_combined}
  \rho=\left[\sum_{i=1}^N\rho_i^2\right]^{1/2}.
\end{equation}
For ET, three arms interfere with each other in pairs, which is equivalent to three independent detectors, thus $N=3$. Generally, if $\rho\geq 8$ we can claim to detect a GW signal.

By matching the GW signals to GW templates we can obtain the luminosity distance to GW source, as well as other parameters. Due to the degeneracy between the luminosity distance $D_L$ and inclination angle $\iota$, the uncertainty on $D_L$ may be very large. However, if the GW event is accompanied by a short gamma-ray burst (GRB, which is expected in the coalescence of NS-NS binary and NS-BH binary), then due to the beaming of GRB outflow we can assume that the inclination angle is small, hence the degeneracy breaks. In this case the uncertainty on $D_L$ can be estimated as \cite{Sathyaprakash:2009xt,Cai:2017sby}
\begin{equation}\label{eq:error_DL}
  \delta D_L^{\rm GW}=\frac{2D_L}{\rho}.
\end{equation}

For GW source at high redshift, there is an additional uncertainty arising from weak lensing effect caused by the intergalactic medium along the line-of-sight. This uncertainty is assume to be proportional to redshift, i.e. $\delta D_L^{\rm lens}/D_L=0.05z$ \cite{Zhao:2011}. Therefore, the total error on $D_L$ is given by
\begin{equation}\label{eq:dL_error}
  \delta D_L=\sqrt{\left(\frac{2D_L}{\rho}\right)^2+(0.05zD_L)^2}\,.
\end{equation}
At low redshift ($z\lesssim 1$), the uncertainty caused by weak lensing is negligible. However, ET can detect GW signals at redshift $z\gtrsim5$. At such high redshift, the uncertainty caused by weak lensing is comparable to the uncertainty caused by detector itself.

\subsection{test the DDR}\label{sec:DDR}

If a GW signal is strongly lensed by a foreground galaxy, we can simultaneously measure the angular diameter distance and luminosity distance to the GW source. The angular diameter distance can be measured from strongly lensing effect according to section {\ref{sec:SL}}, and the luminosity distance can be measured from the GW signals according to section {\ref{sec:GW}}.

One should pay specific attention that, due to the magnification effect of lensing, the luminosity distance measured from the strongly lensed GW signals is not the true distance. From equation (\ref{eq:amplidude}) we know that the luminosity distance $D_L$ is inversely proportional to the amplitude of GW signal, while the latter is magnified by the lensing effect by a factor of $\sqrt{\mu_\pm}$ \cite{Wang:1996}. For the singular isothermal spherical lens, the magnification factor can be calculated as $\mu_\pm =1\pm\theta_E/\beta$, where $\beta$ is the actual position of the source, and ``$\pm$" represent the first and second images, respectively. The actual position of the source $\beta$ can be determined through deep photometric imaging, $\beta/\theta_E=(F_+-F_-)/(F_++F_-)$, where $F_\pm$ are the photometric flux of two images. If the magnification factor is measured from photometric observations, we can obtain the true distance $D_L^{\rm true}=\sqrt{\mu_\pm}D_L^{\rm obs}$. The uncertainty of $\mu_\pm$ will propagate to $D_L$. Therefore, the total uncertainty on $D_L(z_s)$ is given by \cite{{Lin:2020mdh}}
\begin{equation}\label{eq:error_on_dL}
  \frac{\delta D_L^{\rm total}}{D_L}=\sqrt{\left(\frac{2}{\rho}\right)^2+(0.05z_s)^2+\frac{1}{4}\left(\frac{\delta\mu_\pm}{\mu_\pm}\right)^2}.
\end{equation}

Having $D_A$ and $D_L$  measured, we can use them to test DDR. We write the possible deviation of DDR as
\begin{equation}\label{eq:eta_z}
  \frac{(1+z)^2D_A}{D_L}=\eta(z).
\end{equation}
Specifically, we consider two different parametrizations of $\eta(z)$, namely, $\eta_1(z)=1+\eta_0 z$ and $\eta_2(z)=1+\eta_0 z/(1+z)$. The parameter $\eta_0$ represents the amplitude of deviation from the standard DDR. If $\eta_0=0$, the standard DDR holds. By fitting the measured $D_A$ and $D_L$ to equation (\ref{eq:eta_z}), $\eta_0$ can be constrained. The best-fitting $\eta_0$ can be obtained by maximizing the following likelihood,
\begin{equation}
  \mathcal{L}\propto\prod_{i=1}^N\frac{1}{\sqrt{2\pi}\sigma_{\rm total}}\exp\left[-\left(\frac{(1+z)^2D_A-\eta(z)D_L}{\sigma_{\rm total}}\right)^2\right],
\end{equation}
where
\begin{equation}
  \sigma_{\rm total}=\sqrt{(1+z)^4(\delta D_A)^2+\eta^2(z)(\delta D_L)^2},
\end{equation}
and the product runs over all the data points.

\section{Monte Carlo simulations}\label{sec:simulations}

In this section, based on the designed sensitivity of ET, we will use Monte Carlo simulations to investigate the precision of strongly lensed GWs in constraining DDR. The fiducial cosmological model is chosen to be the flat $\Lambda$CDM model, with parameters $\Omega_m=0.3$ and $H_0=70~{\rm km~s}^{-1}~{\rm Mpc}^{-1}$. The luminosity distance in the fiducial cosmological model is given by
\begin{equation}\label{eq:dL}
  \bar{D}_L=(1+z)\frac{c}{H_0}\int_0^z\frac{dz}{\sqrt{\Omega_m(1+z)^3+1-\Omega_m}}.
\end{equation}

\begin{figure}[htbp]
\centering
\includegraphics[width=0.5\textwidth]{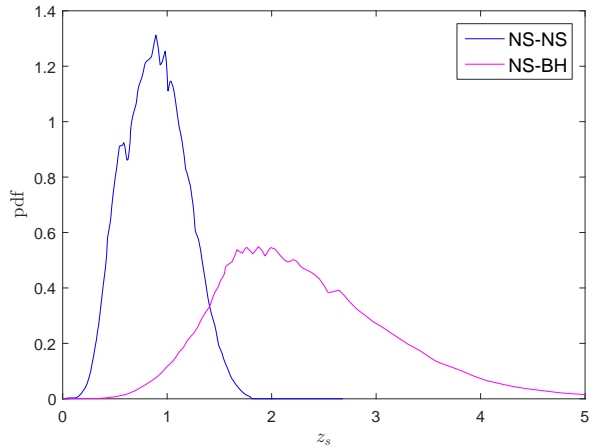}
\caption{\label{fig:redshift_zs} The redshift distribution of strongly lensed GW sources. The lines are reproduced from Ref.\cite{Biesiada:2014kwa}, but are renormalized such that the area under each line is unity.}
\end{figure}

We only consider the GWs produced by the coalescence of NS-NS and NS-BH binaries. BH-BH binaries are not considered, because according to most theoretical models the coalescence of BH-BH binary has no electromagnetic counterparts, although some exotic models predict that it may also be accompanied by electromagnetic counterparts \cite{Zhang:2016rli,Fraschetti:2016bpm,Perna:2016jqh}. The redshift distribution and event rate of GWs depend on the stellar evolution model. Ref.\cite{Biesiada:2014kwa} has calculated in detail the redshift distribution and event rate of the strongly lensed inspiral double compact objects (including NS-NS, NS-BH and BH-BH binaries) in different scenarios. Based on the initial configuration of ET, the excepted redshift distribution of the strongly lensed NS-NS and NS-BH events in the standard evolution scenario is plotted in Figure \ref{fig:redshift_zs}.

The probability density function (pdf) of a GW event at redshift $z_s$ being lensed by a foreground galaxy at redshift $z_l$ $(z_l<z_s)$ is given by \cite{Biesiada:2014kwa}
\begin{equation}\label{eq:pdf_of_zs}
  P(z_l|z_s)=C\frac{\tilde{r}^2(z_l,z_s)\tilde{r}^2(0,z_l)}{\tilde{r}^2(0,z_s)E(z_l)},
\end{equation}
where $E(z)=\sqrt{\Omega_m(1+z)^3+1-\Omega_m}$ is the dimensionless Hubble parameter, $C$ is a normalization constant, and
\begin{equation}
  \tilde{r}(z_1,z_2)=\int_{z_1}^{z_2}\frac{1}{E(z)}dz
\end{equation}
is the dimensionless comoving distance from $z_1$ to $z_2$. For a given GW source at redshift $z_s$, the redshift of the lens galaxy $z_l$ is randomly sampled according to the pdf given in equation (\ref{eq:pdf_of_zs}). We assume that $z_s$ and $z_l$ can be measured spectroscopically, so the uncertainty is negligible.

The velocity dispersion of the lens galaxy is assumed to follow the modified Schechter function \cite{Choi:2007}
\begin{equation}
  n(\sigma)d\sigma=n_0\left(\frac{\sigma}{\sigma_*}\right)^\alpha\exp\left[-\left(\frac{\sigma}{\sigma_*}\right)^\beta\right]\frac{d\sigma}{\sigma},
\end{equation}
where $n_0$ is a normalization constant, $\sigma_*=161~{\rm km~s}^{-1}$, $\alpha=2.32$ and $\beta=2.67$. We set a lower limit on the velocity dispersion, i.e., $\sigma_{\rm lower}=70~{\rm km~s}^{-1}$. The observational accuracy of velocity dispersion may strongly affect the accuracy of $D_A$. According to the presently available quasar lensing systems compiled in Ref.\cite{Cao:2015qja}, the measured uncertainty of velocity dispersion is about $10\%$. With the progress of observational technique, it is not impossible to reduce the uncertainty to less than $5\%$ in the near future \cite{Cao:2019kgn}.

To determine $D_A$, it is necessary to precisely measure the Fermat potential difference $\Delta\phi$, the Einstein angle $\theta_E$, and the time delay between two images $\Delta t$. Benefitting from the fact that GW signals do not suffer from the bright AGN contamination from the lens galaxy, the measured accuracy of $\Delta\phi$ can be improved to $\sim 0.6\%$ in the lensed GW system, while the uncertainty in the lensed quasar systems is approximately larger by a factor of five \cite{Liao:2017ioi}. The accuracy of $\theta_E$ is expected to be at $\sim 1\%$ level in the future LSST survey \cite{Cao:2019kgn}. Thanks to the transient property of GW events, the arrival time of GW signals can be accurately recorded, so the uncertainty on time delay is negligible.

To correctly determine $D_L$, we should precisely measure the magnification factor $\mu_\pm$ from photometric observations. In general, $\mu_\pm$ can be determined by measuring the photo flux of two images as described in section {\ref{sec:DDR}}. The measurement of image flux, however, may be highly uncertain due to the photometric contamination from the foreground lensing galaxy. Here we follow Ref.\cite{Cao:2019kgn} and assume a $\sim 20\%$ uncertainty on $\mu_\pm$.

The masses of the neutron star and of the black hole are assumed to be uniformly distributed in the range $m_{NS}\in U(1,2)M_\odot$ and $m_{BH}\in U(3,10)M_\odot$, respectively \cite{Cao:2019kgn}. In addition, we assume that the GW sources are uniformly distributed in the sky, i.e., $\theta\in U(0,\pi)$, $\varphi\in U(0,2\pi)$. The occurrence rate of lensed NS-NS and NS-BH events depends on the stellar evolution scenarios. Numerical simulations show that the NS-BH event rate is in general larger than the NS-NS event rate \cite{Biesiada:2014kwa}. In our simulations, we assume that the ratio of NS-NS event rate and NS-BH event rate is $1:5$.

Based on the discussions above, we assume that the measured accuracy for parameters ($\sigma,\theta_E,\Delta\phi,\mu_\pm$) is ($5\%,1\%,0.6\%,20\%$), respectively. With this setup, we can now simulate the strongly lensed GW events following the steps bellow:
\begin{enumerate}
  \item Randomly sample parameters $(z_s, z_l, \sigma, m_1, m_2, \theta, \varphi)$ according to the pdf of each parameter described above.
  \item Calculate the SNR of GW signal based on the simulated parameters according to section {\ref{sec:GW}}. If ${\rm SNR}>16$, continue; else go back to step 1.
  \item Calculate $D_A$ and $\delta D_A$ at redshift $z_s$ according to section {\ref{sec:SL}}. If $\delta D_A/D_A < 30\%$, continue; else go back to step 1.
  \item Calculate the fiducial luminosity distance $\bar{D}_L$ at redshift $z_s$ according to equation (\ref{eq:dL}), and the uncertainty $\delta D_L$ according to equation (\ref{eq:error_on_dL}).
  \item Sample $D_L$ from the Gaussian distribution $D_L\sim G(\bar{D}_L,\delta D_L)$.
  \item Save the parameter set $(z_s,D_A,\delta D_A,D_L,\delta D_L)$ as an effective GW event; go back to step 1 until we obtain $N$ events.
\end{enumerate}

Some notes on the simulation procedures. In step 2, we require that the SNR of GW signal is at least ${\rm SNR}\gtrsim 16$, compared to the usual criterion ${\rm SNR}\gtrsim 8$. For a GW events with ${\rm SNR}\approx 8$, the uncertainty on $D_L$ from GW signal itself is about 25\%. If the errors from weak lensing and magnification factor are included, the total uncertainty on $D_L$ is $\sim 30\%$ for an event at $z_s=2$, which is unacceptably large. If we require ${\rm SNR}>16$, the uncertainty can be reduced down to $20\%$. In step 3, we only retain the GW events whose accuracy on $D_A$ are better than $30\%$. From equation (\ref{eq:error_DA_zs}) we can see that the error on $D_A$ mainly comes from the error on $R_A$, while the latter is at the order of $10\%$. The error on $D_A$ is larger than the error on $R_A$ by a factor of  $1/(1-R_A)$. If $R_A$ is close to unity (this happens when $z_l\ll z_s$), the error on $D_A$ may be very large.

\begin{figure}[htbp]
\centering
\includegraphics[width=0.42\textwidth]{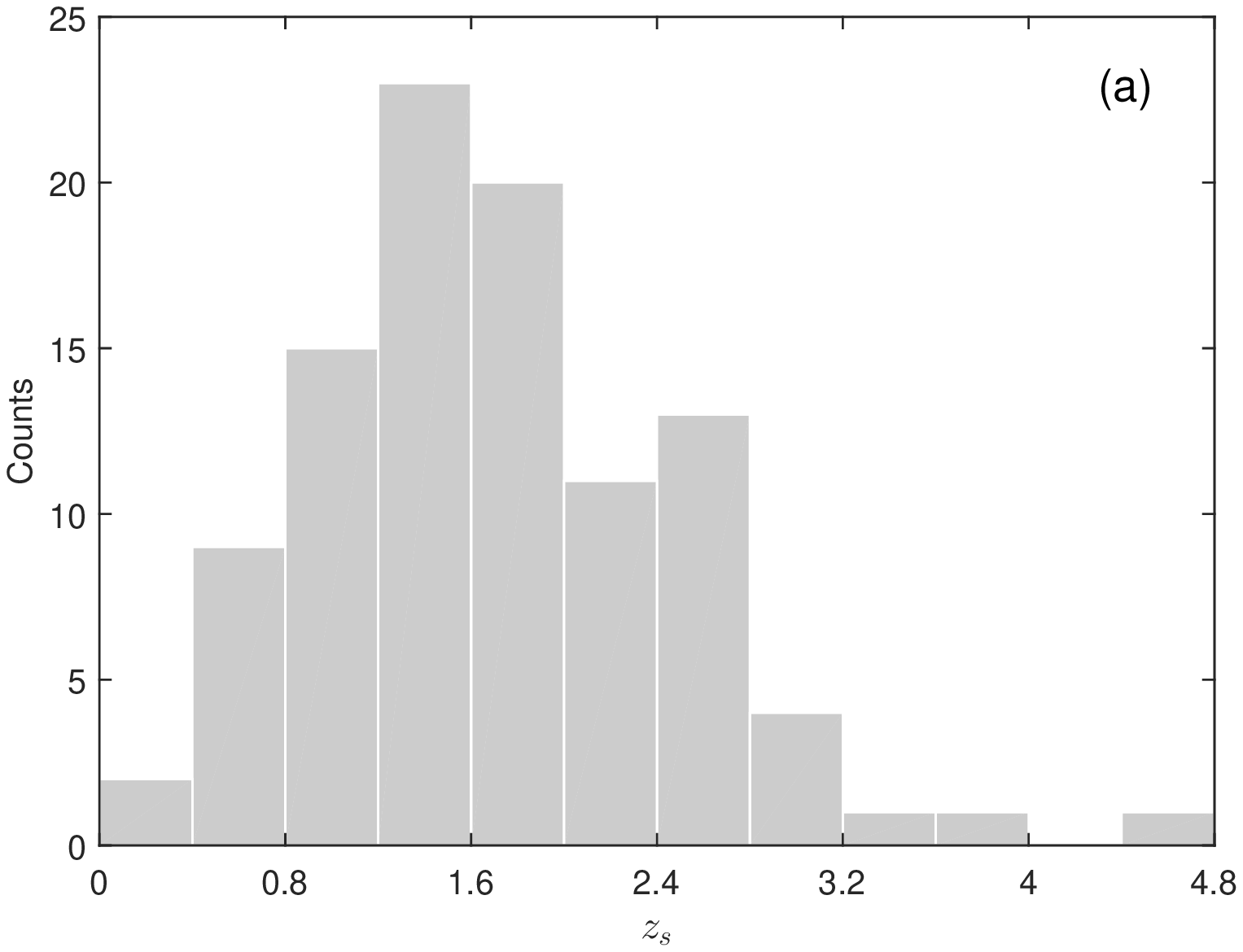}
\includegraphics[width=0.42\textwidth]{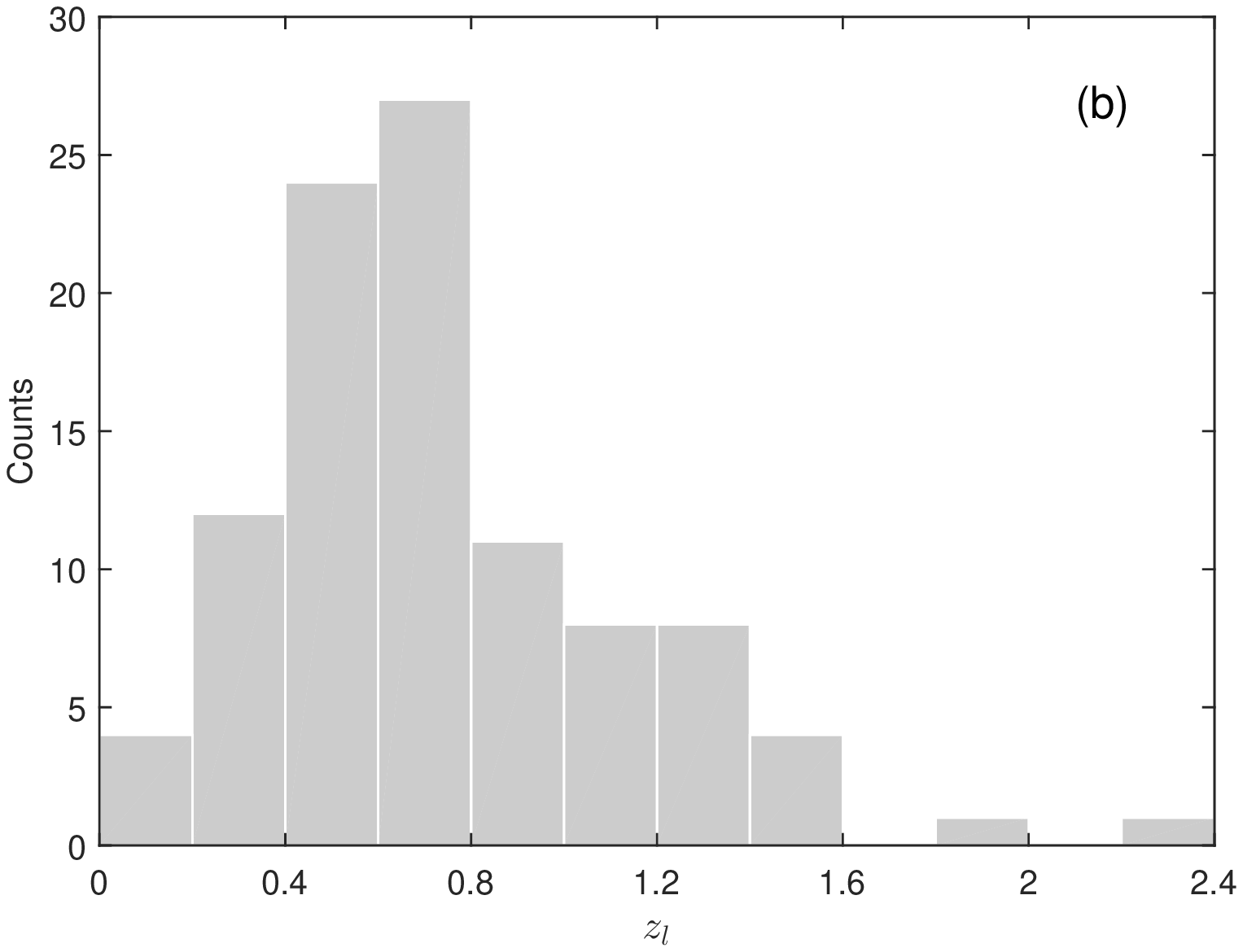}
\includegraphics[width=0.42\textwidth]{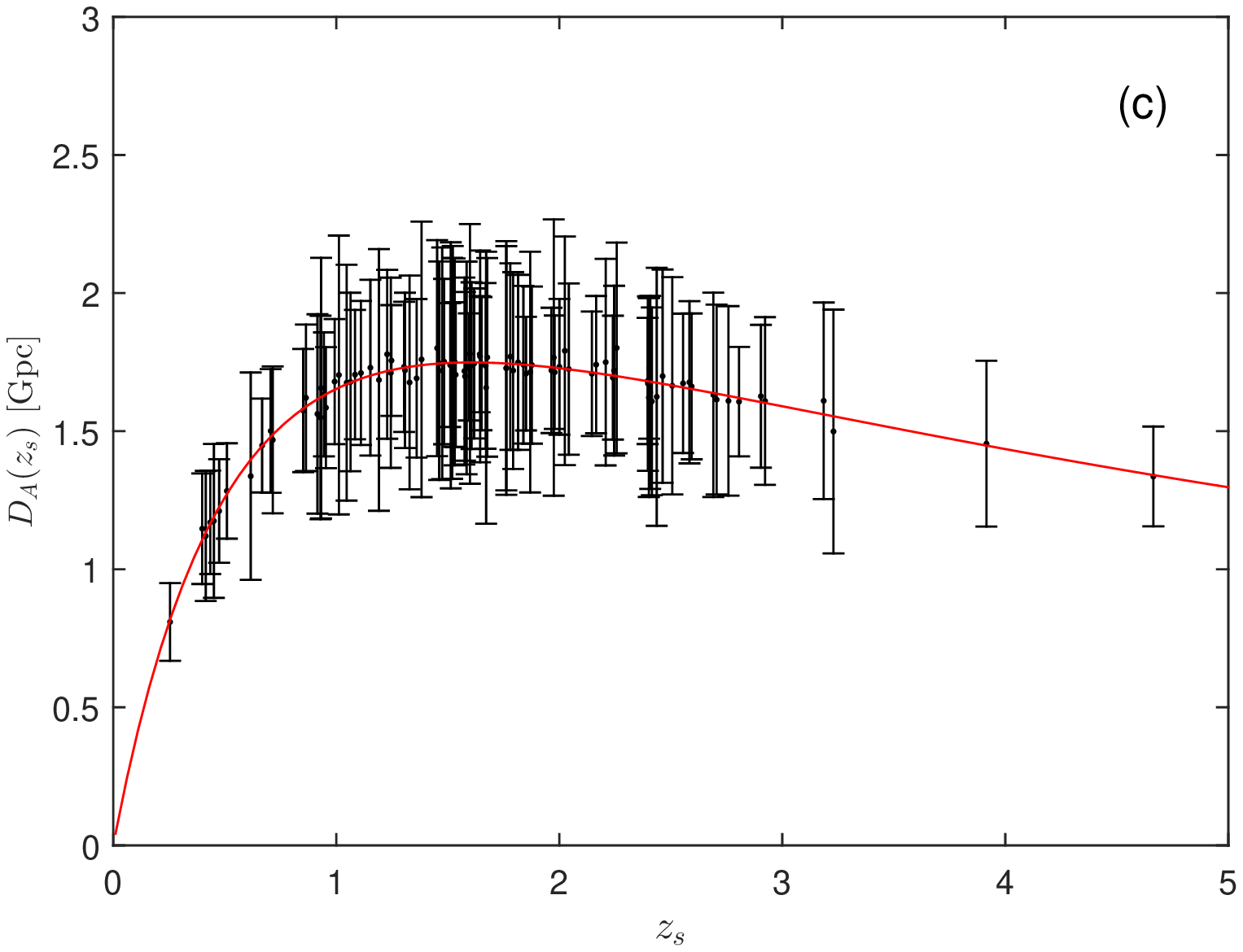}
\includegraphics[width=0.42\textwidth]{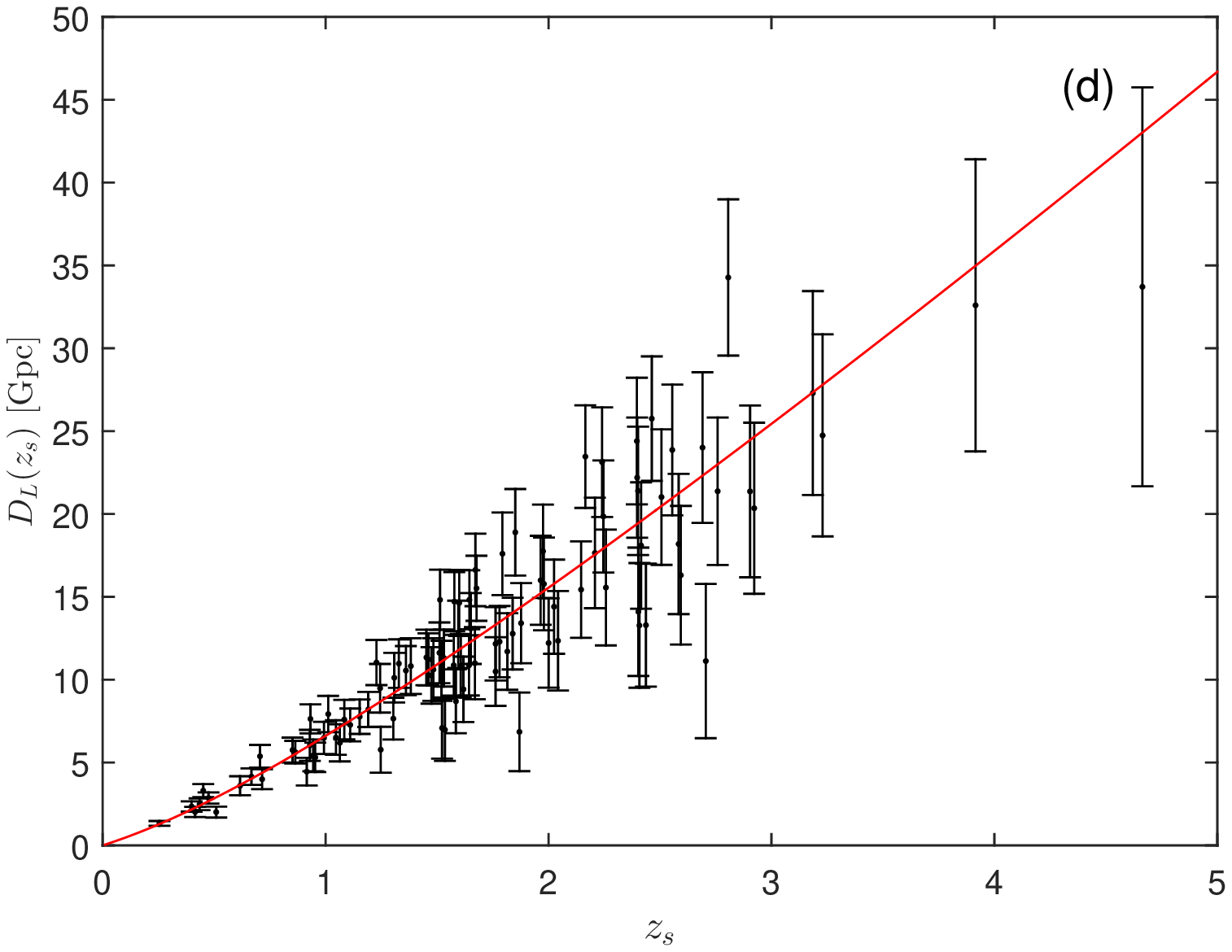}
\caption{\label{fig:results} A representative simulating result of 100 strongly lensed events. (a) The redshift distribution of $z_s$. (b) The redshift distribution of $z_l$. (c) The angular diameter distance. (d) The luminosity distance. The red lines are the theoretical curves of the fiducial $\Lambda$CDM model.}
\end{figure}

A representative simulating result of 100 strongly lensed events is plotted in Figure \ref{fig:results}. Panel (a) and panel (b) show the histogram of $z_s$ and $z_l$, respectively. Panel (c) and panel (d) show the angular diameter distance and luminosity distance versus redshift $z_s$, respectively. The red lines are the theoretical curves of the fiducial $\Lambda$CDM model. The redshift distributions of GW source and lens galaxy peak at about 1.6 and 0.6, respectively. The uncertainty on $D_L$ increases with redshift because of the increasing error caused by the weak lensing effect at high redshift.


\begin{figure}[htbp]
\centering
\includegraphics[width=0.5\textwidth]{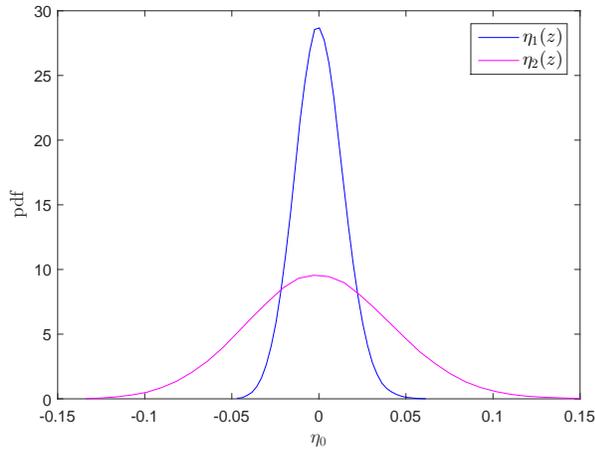}
\caption{\label{fig:dpf} The posterior pdf of $\eta_0$ constrained by 100 strongly lensed GW events.}
\end{figure}

Using the simulated GW events, DDR can be strictly constrained. The posterior pdf of $\eta_0$ constrained from 100 simulated GW events is plotted in Figure \ref{fig:dpf}. With 100 strongly lensed GW events, the parameter $\eta_0$ can be constrained at $\sim 1.3\%$ and $\sim 3\%$ levels for the first and second parametrizations, respectively. The simulating results imply that strongly lensed GW events are very promising in constraining DDR as the construction of ET in the future.

\section{Discussion and conclusions}\label{sec:discussion}

In this paper, we have investigate the possibility of using strongly lensed GW event to constrain DDR. The strongly lensed GW events provide a unique way to measure angular diameter distance and luminosity distance to the GW source simultaneously, thus can be directly used to test DDR. This method is independent of cosmological model, except the assumption that the universe is spatially flat. Monte Carlo simulations shows that with about 100 such events, DDR can be constrained at $\sim 1.3\%$ and $\sim 3\%$ level for the first and second parametrizations, respectively. In comparison, using the combination of SNe, galaxy clusters and BAO data, DDR is constrained at $\sim 12\%$ and $\sim 22\%$ level for the first and second parametrizations, respectively \cite{Lin:2018mdj}. Using the combination of SNe and ultra-compact radio sources, DDR is constrained at $\sim 5\%$ and $\sim 16\%$ level for the first and second parametrizations, respectively \cite{Li:2018}. Ref.\cite{Liao:2019xug} have used the combined data of GWs and strongly lensed quasar systems to constrain DDR, and the constraining accuracy is comparable to our results. Note that the method we proposed here is completely different from that in Ref.\cite{Liao:2019xug}. In Ref.\cite{Liao:2019xug}, the strongly lensed quasars provide $D_A$, and GW events provide $D_L$. So $D_A$ and $D_L$ are still measured from different source at different redshift. While the method we proposed here measures $D_A$ and $D_L$ from the same GW source.

The biggest challenge to put the method into practise is how to identify the strongly lensed GW events. The angular separation between two images of a typical strong lensing system is of the order of arc seconds. It seems extremely difficult to reach such high angular resolution in the near future. Zhao \& Wen \cite{Zhao:2018cbb} found that even for a network of three or four third generation GW detectors, the localization accuracy is about several arc degrees. This accuracy is far from enough to separate the images, but it is enough to identify the host galaxy. If two GW signals with the same observed strains (up to a normalization constant) come from the same host galaxy, and if the relative time delay is consistent with theoretical prediction, then these two signals can be treated as two images of a strongly lensed GW event. If we further assume that GW and electromagnetic wave travel along the same null-geodesic, the image separation between two GW signals can be obtained through photometric observations, which can be easily realized with the present technique.

The method proposed here needs independent measurement of redshift of source and lens. The third general GW detector ET is expected to be able to record several hundreds strongly lensed GW events during its lifetime \cite{Biesiada:2014kwa}. Unfortunately, most of the events are produced by the coalescence of BH-BH binaries, which is in general has no electromagnetic counterparts. Without electromagnetic counterparts, it is difficult to identify the host galaxy of GW sources, hence it is impossible to measure the spectroscopic redshift. This prevents the direct use of these events to test DDR. However, if the GW event can be precisely localized, it is possible to infer the redshift of GW source statistically \cite{Chen:2017rfc,Mukherjee:2020hyn}. Of course, this will introduce additional uncertainty.

\acknowledgments{This work has been supported by the National Natural Science Fund of China under grant Nos. 11603005, 11775038 and 11947406.}

\end{document}